\documentclass[twocolumns,bibyear]{aa} % for a paper on 1 column  
\usepackage{graphicx,epstopdf}
\usepackage{url}
\usepackage{txfonts}
\usepackage{amsmath}
\def\Q{\ifmmode\mathcal{Q}\else$\mathcal{Q}$\fi}

\begin{document} 

   \title{Stellar collisions in globular clusters: Constraints on the initial mass function of the first generation of stars}

   \author{Sami Dib\inst{1}, Valery V. Kravtsov\inst{2}, Hosein Haghi\inst{3}, Akram Hasani Zonoozi\inst{3}, Jos\'{e} Antonio Belinch\'{o}n\inst{4}}
 
   \institute{Max Planck Institute for Astronomy, K\"{o}nigstuhl 17, 69117, Heidelberg, Germany\\
      \email{sami.dib@gmail.com; dib@mpia.de}
        \and       
             Sternberg Astronomical Institute, Lomonosov Moscow State University, University Avenue 13, 119899 Moscow, Russia 
       \and 
              Department of Physics, Institute for Advanced Studies in Basic Sciences (IASBS), Zanjan 45137-66731, Iran
       \and 
              Departamento de Matem\'{a}ticas, Universidad de Atacama, Av. Copayapu 485, Copiap\'{o}, Chile
                      }
          
\authorrunning{Dib et al.}
\titlerunning{Stellar collisions in globular clusters and constraints on the IMF}
         
% \abstract{}{}{}{}{} 
% 5 {} token are mandatory
 
\abstract{Globular clusters display an anticorrelation between the fraction of the first generation of stars ($N({\rm G1})/N({\rm tot})$) and the slope of the present-day mass function of the clusters ($\alpha_{pd}$), which is particularly significant for massive clusters. In the framework of the binary-mediated collision scenario for the formation of the second-generation stars in globular clusters, we test the effect of a varying stellar initial mass function (IMF) of the G1 stars on the $(N({\rm G1})/N({\rm tot}))-\alpha_{pd}$ anticorrelation. We use a simple collision model that has only two input parameters, the shape of the IMF of G1 stars and the fraction of G1 stars that coalesce to form second-generation stars. We show that a variable efficiency of the collision process is necessary in order to explain the $(N({\rm G1})/N({\rm tot}))-\alpha_{pd}$ anticorrelation; however, the scatter in the anticorrelation can only be explained by variations in the IMF, and in particular by variations in the slope in the mass interval $\approx$ (0.1-0.5) M$_{\odot}$. Our results indicate that in order to explain the scatter in the $(N({\rm G1})/N({\rm tot}))-\alpha_{pd}$ relation, it is necessary to invoke variations in the slope in this mass range between $\approx -0.9$ and $\approx -1.9$. Interpreted in terms of a Kroupa-like broken power law, this translates into variations in the mean mass of between $\approx 0.2$ and $0.55$ M$_{\odot}$. This level of variation is consistent with what is observed for young stellar clusters in the Milky Way and may reflect variations in the physical conditions of the globular cluster progenitor clouds at the time the G1 population formed or may indicate the occurrence of collisions between protostellar embryos before stars settle on the main sequence.}


 \maketitle

%
%-------------------------------------------------------------------

\section{Introduction}\label{introduction}

The stellar initial mass function (IMF) of stars (i.e., the distribution of the masses of stars at their birth) is of fundamental importance in astrophysics. The IMF controls the efficiency of star formation in molecular clouds (Dib et al. 2011,2013; Hony et al. 2015), the radiative and mechanical feedback from stars into the large-scale interstellar medium (Dib et al. 2006,2021; Martizzi et al. 2016; Silich \& Tenorio-Tagle 2017), and the dynamical and chemical evolution of galaxies (C\^{o}t\'{e} et al. 2016). Significant efforts have been devoted to the determination of the shape of the IMF in a variety of environments, including the Galactic field (Salpeter 1955; Bochanski et al. 2010; Rybizki \& Just 2015; Mor et al. 2019; Sollima 2019), the Galactic bulge (Zoccali et al. 2000; Wegg et al. 2017), and globular clusters (GCs; Da Costa \& Freeman 1976; De Marchi \& Paresce 1997; Covino \& Ortolani 1997; Marconi et al. 1998; Paresce \& De Marchi 2000; Sollima et al. 2007; Balbinot et al. 2009; Sollima \& Baumgardt 2017; Cadelano et al. 2020), as well as in many young open clusters and associations (Dib 2014; Weisz et al. 2015; Maia et al. 2016; Dib et al. 2017; Jose et al. 2017; Madaan et al. 2020; Bisht et al. 2021; Damian et al. 2021; Elsanhoury et al. 2022). The existence of IMF variations in clusters has significant consequences for the galaxy-wide IMF and for galactic evolution (Dib \& Basu 2018; Dib 2022). In the galactic ecosystem, GCs stand out as relics of early star formation with ages that span the range $\approx 11$ Gyr to $\approx 13.2$ Gyr (Salaris et al. 1997; VandenBerg et al. 2013; Pfeffer et al. 2018; Usher et al. 2019; Oliveira et al. 2020), and it is well established that GCs in the Milky Way and in the Magellanic clouds harbor two or more stellar populations (d'Antona \& Caloi 2008; Milone et al. 2010; Sbordone et al. 2011; Gratton et al. 2012; Cummings et al. 2014; Piotto et al. 2015; Lee 2015; Oldham \& Auger 2016; Mucciarelli et al. 2016; Massari et al. 2016; Dalessandro et al. 2016; Bowman et al. 2017; Carretta \& Bragaglia 2018; Latour et al. 2019; Gilligan et al. 2020; Dondoglio et al. 2021; Jang et al. 2021; D'Antona et al. 2022; Kapse et al. 2022). 

The origin of the multiple stellar populations in GCs is highly debated. A number of scenarios have been proposed to explain the existence of two or more stellar populations. The oldest and most popular of these is the asymptotic giant branch (AGB) self-enrichment scenario, which is based on the formation of second-generation stars (hereafter G2 stars) from the ejecta of the first-generation (G1) AGB stars after gas has been cleared from the cluster by feedback from massive stars (Cottrell \& Da Costa 1981; D'Ercole et al. 2016; Bekki 2017). Since the pure AGB scenario fails to reproduce the observed Na-O anticorrelation, a modified version of the model was introduced whereby pristine gas is accreted from outside the cluster and mixes with the AGB ejecta to form the G2 stars (Ventura \& D'Antona 2009; Calura et al. 2019; Yaghoobi et al. 2022). Other variants of the model relied on the ejecta of fast rotating massive stars (FRMSs) instead of AGB stars (Decressin et 2007; Krause et al. 2013) or on the rapid efficient cooling of stellar wind material that falls back to the center of the cluster and mixes with leftover gas with a G1-like chemical composition (W\"{u}nsch et al. 2017). Other models invoked a dichotomy in GC mass whereby the efficiency of the AGB ejecta retention in the cluster would depend on the cluster mass (Valcarce \& Catelan 2011). The AGB- and FRMS-based scenarios suffer from the same types of shortcomings. It is unclear whether gas ejecta of AGB stars (or FRMSs) can cool and form the second generation of stars due to uninterrupted heating by G1 stars, particularly from X-ray binaries, that heat the gas and prevent it from fragmenting (Conroy \& Spergel 2011). Another issue is the so-called mass-budget problem. This problem reflects the inability of G1 stars, under the assumption of a Milky Way-like IMF, to produce enough chemically enriched material for the formation of significant populations of G2 stars. Furthermore, and owing to the GCs orbits in the galaxy, it is not clear whether accretion of external gas would be efficient (Conroy \& Spergel 2011). Khalaj \& Baumgardt (2015) investigated under which conditions the loss of G1 stars due to gas expulsion from the protocluster cloud can help reproduce the fraction of G2 stars and some of the clusters properties, such as their radial profiles and mass-half mass-radius relation. They find that reproducing the observations requires rapid gas expulsion timescales ($\lesssim 10^{5}$ yr) and a large number of supernova explosions, three to six times more than what would be found in a Galactic field-like IMF.

Another physical process that can lead to the formation of G2 stars over a short timescale, prior to gas expulsion from the clusters by the first supernovae, is stellar collisions (Sills et al. 2002; Sills \& Glebbeek 2010; Jiang et al. 2014; Wang et al. 2020). Collisions can still occur at later times, albeit at a lower rate (Kravtsov \& Calder\'{o}n 2021; Kravtsov et al. 2022). Collisions not only set the relative fractions of the G1 and G2 stars, but they also alter the shape of the IMF of the G1 stars (Dib et al. 2007; Kravtsov et al. 2022). While more work is still needed in order to understand whether stellar collisions, particularly of low-mass stars, can lead to the chemical anticorrelations that are observed for GCs, Kravtsov et al. (2022) present observational evidence that supports the scenario in which stellar collisions of G1 stars in the mass range $\approx$ (0.1-0.5) M$_{\odot}$ can lead to the formation of G2 stars in the mass range (0.5-0.9) M$_{\odot}$. In particular, they presented evidence for the existence of an anticorrelation between the fraction of G1 stars ($N({\rm G1})/N({\rm tot})$) and the slope of the present-day mass function of GCs in the stellar mass range (0.2-0.8) M$_{\odot}$ ($\alpha_{pd}$). They also find that the fraction of G1 stars is anticorrelated with the present-day encounter rate measured in these clusters. The present-day encounter rates of the clusters are expected to scale with their primordial values when the clusters were denser (e.g., Maccarone \& Peacock 2011). Using a simple collision model with a parametrized collision efficiency and a Kroupa-like IMF for the G1 stars (Kroupa 2001), Kravtsov et al. (2022) showed that it is possible to reproduce the $(N({\rm G1})/N({\rm tot}))-\alpha_{pd}$ anticorrelation both in terms of slope and absolute values. Nonetheless, the $(N({\rm G1})/N({\rm tot}))-\alpha_{pd}$ anticorrelation shows a level of scatter that cannot be simply explained by a fixed IMF nor by observational uncertainties. In this paper, and in the framework of this collision-based scenario, we focus on the role of the IMF in explaining both the trend and scatter in the $(N({\rm G1})/N({\rm tot}))-\alpha_{pd}$ relation. In particular, we explore how variations in the shape of the IMF can impact the shape of the $(N({\rm G1})/N({\rm tot}))-\alpha_{pd}$ anticorrelation. In Sects. \ref{dat} and \ref{mod} we briefly recall the observational data and the main elements of the model that is used to reproduce them, respectively. The results are presented in Sect. \ref{res}, and in Sect. \ref{conc} we conclude.

\section{Data}\label{dat}

We used the sample of multiple stellar populations in Galactic GCs obtained by Milone et al. (2017). Using uniform multiband \textit{Hubble} Space Telescope (HST) photometry (Sarajedini et al. 2007; Piotto et al. 2015) in the central parts of a sample of 57 GCs, Milone et al. (2017) isolated red giant branch (RGB) stars and measured the fractions of the G1 population to the total number of RGB stars ($N({\rm G1})/N({\rm tot})$). These fractions were measured for a total of 54 GCs. For the slope of the present-day mass function, we used the compilation of Ebrahimi et al. (2020), which includes 32 GCs and for which the slope was measured by fitting a power law in the stellar mass range (0.2-0.8) M$_{\odot}$. The cluster NGC 6584 was excluded from our sample since the data for this GC are incomplete. The data for four additional GCs were taken from Sollima \& Baumgardt (2017), namely NGC 4833, NGC 6205, NGC 6397, and NGC 6656, and the measurement of $\alpha_{pd}$ for these clusters was also performed by fitting a single power law over the same stellar mass range of (0.2-0.8) M$_{\odot}$. In total, we have a sample of 35 GCs with measurements of $(N({\rm G1})/N({\rm tot}))$ and $\alpha_{pd}$.

It is worth pointing out that the measurement of the slope of the present-day mass function by Ebrahimi et al. (2020) relies on the entire stellar census of the clusters. On the other hand, the fractions of G1 stars measured by Milone et al. (2017) were obtained within the HST/WFC3 (Wide Field Camera 3) field of view, and as such, these measurements are representative of the fractions in the central regions of the clusters. The fractions of G1 stars measured by Milone et al. (2017) underestimate their true values when measured for the entire clusters since G2 stars are often observed to be more concentrated toward the clusters' centers (e.g., Carretta et al. 2009; Kravtsov \& Calder\'{o}n 2021; see, however, Dalessandro et al. 2014 for a counterexample). Thus, the true values of the G1 star fractions are expected to be, on average, systematically higher than the values estimated by Milone et al. (2017).

\section{Model}\label{mod}

In the framework of the collision-based scenario, our aim is to understand the effect of the shape of the IMF of the G1 stars on the observed $(N({\rm G1})/N({\rm tot}))-\alpha_{pd}$ anticorrelation. Understanding the precise effects of collisions requires calculating the collision rates over the entire stellar mass range and a knowledge of the coalescence efficiency as a function of stellar mass (e.g., Dib et al. 2007). Here, we used a simpler model with two mass bins, similar to the one used in Kravtsov et al. (2022). We briefly recall its basic elements. We assume that stars that can coalesce are those exclusively found in the stellar mass range (0.1-0.5) M$_{\odot}$. The collision products (G2 stars) are more massive stars, with masses that fall in the range (0.5-0.9) M$_{\odot}$. If $N_{1}({\rm G1})$ and $N_{2}({\rm G1})$ are the total numbers of G1 stars that fall in the target and product mass bins, respectively, then the fractions of these two population of stars to the total number of stars ($N^{'}_{tot}$) can be written as

\begin{equation}
f_{1,in}=\frac{N_{1}({\rm G1})}{N^{'}_{tot}}
\label{eq1}
\end{equation} 

and

\begin{equation}
f_{2,in}=\frac{N_{2}({\rm G1})}{N^{'}_{tot}}.
\label{eq2}
\end{equation}

If $N_{1,ext}({\rm G1})=f_{ext} N_{1}({\rm G1})$ is the total number of stars that have experienced a collision, under the assumption that we can only have two stars colliding to form a new one, and with $f_{ext}$ being the fraction of $N_{1}$(G1) stars that undergo a collision, then the new number of stars becomes 

\begin{equation}
N_{tot}=N^{'}_{tot}-\frac{f_{ext} N_{1}({\rm G1})}{2}.
\label{eq3}
\end{equation}

After the collision process, a second generation of stars is formed in the mass range (0.5-0.9) M$_{\odot}$ (i.e., G2 stars). The new fractions of stars in the mass ranges mentioned above will be given by

\begin{equation}
f_{1,fl}=\frac{N_{1}({\rm G1})-f_{ext}N_{1}({\rm G1})}{N_{tot}}=\frac{2 f_{1,in} (1-f_{ext})}{2-(f_{ext}f_{1,in})}
\label{eq4}
\end{equation}

and

\begin{equation}
f_{2,fl}=\frac{N_{2}({\rm G1})+f_{ext}N_{1}({\rm G1})/2}{N_{tot}}=\frac{2 f_{2,in}+f_{ext} f_{1,in}}{2-(f_{ext}f_{1,in})}.
\label{eq5}
\end{equation}The value of $(N({\rm G1})/N({\rm tot}))$ for RGB stars is given by its analog for main sequence (MS) stars as

\begin{equation}
g_{1}=\frac{N_{2}({\rm G1})}{N_{2}({\rm G1})+N_{2}({\rm G2})}=\frac{2 f_{2,in}}{2 f_{2,in}+f_{ext} f_{1,in}}.
\label{eq6}
\end{equation}

For the GCs in our sample, we have access to $(N({\rm G1})/N({\rm tot}))$ and $\alpha_{pd}$. In the model, and for any given functional form of the IMF, we can measure the values of $f_{1,in}$ and $f_{2,in}$, and for a given value of $f_{ext}$ we can calculate the values of $f_{1,fl}$, $f_{2,fl}$, and $g_{1}$. With $f_{1,fl}$ and $f_{2,fl}$ we can also derive the value of the post-collision value of the slope, $\alpha_{pc}$. We calculated the value of $\alpha_{pc}$ as being $\alpha_{pc}=({\rm log}(f_{1,fl})-{\rm log}(f_{2,fl}))/({\rm log}(0.7)-{\rm log}(0.3))$, where $0.3$ and $0.7$ (in units of M$_{\odot}$) are the mean masses in the intervals (0.1-0.5) M$_{\odot}$ and (0.5-0.9) M$_{\odot}$, respectively. We note that the parameter $g_{1}$ for MS stars is formally equivalent to the ratio $(N({\rm G1})/N({\rm tot}))$ deduced from the observations of RGB stars. However, some caution should be exercised. As already noted in Kravtsov et al. (2022) and in Sect. \ref{dat}, the fractions $(N({\rm G1})/N({\rm tot}))$ have been measured in the central parts of GCs. Since RGB G2 stars are typically observed to be more centrally located than their G1 counterparts, the real $(N({\rm G1})/N({\rm tot}))$ fractions should be systematically higher than the values that are actually derived. In contrast, the parameter $g_{1}$ in the models is assumed to be measured for the clusters as a whole. Furthermore, $g_{1}$ is assumed to be a proxy for MS stars that are now found on the RGB. However, the observed fraction $(N({\rm G1})/N({\rm tot}))$ could also contain, in addition to RGB stars whose masses fall in the range $\approx$ (0.8-0.85) M$_{\odot}$, a contribution from more massive blue straggler stars that are either of collisional origin or have been formed by mass transfer in binaries. For the reasons discussed above, we prefer to formally distinguish between $(N({\rm G1})/N({\rm tot})),$ which is derived from the observations, and $g_{1}$, which is calculated in the models, even if in practice the two quantities are expected to be very close. 
  
\begin{figure}
\begin{center}
\includegraphics[width=\columnwidth] {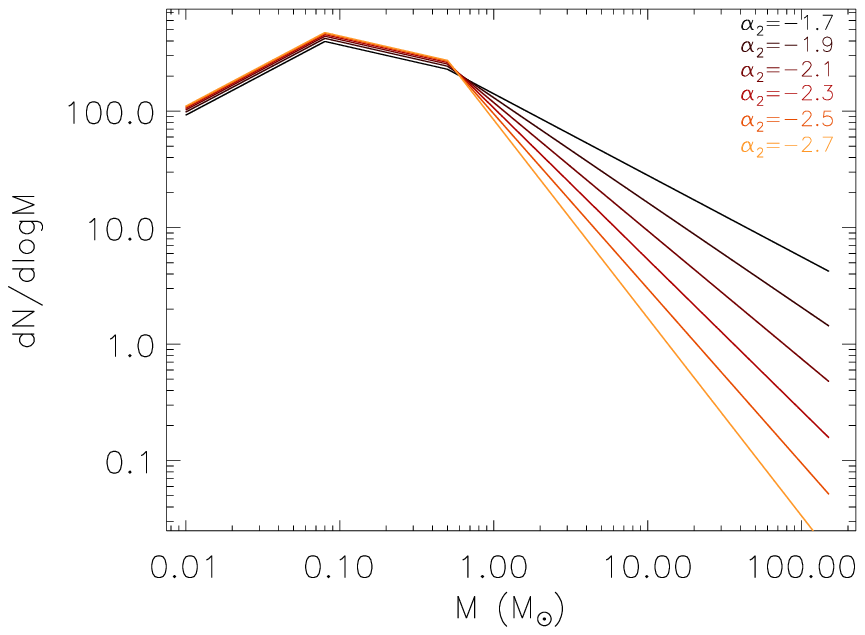}
\caption{Shape of the IMF for various values of the slope at the high mass end, $\alpha_{2}$, around the Galactic field value of $-2.3$. The values of $\alpha_{0}$ and $\alpha_{1}$ are set to the Galactic field values of $-0.3$ and $-1.3$, respectively. The mass functions are normalized assuming the same arbitrary mass.}
\label{fig1}
\end{center}
\end{figure}

\begin{figure}
\begin{center}
\includegraphics[width=\columnwidth] {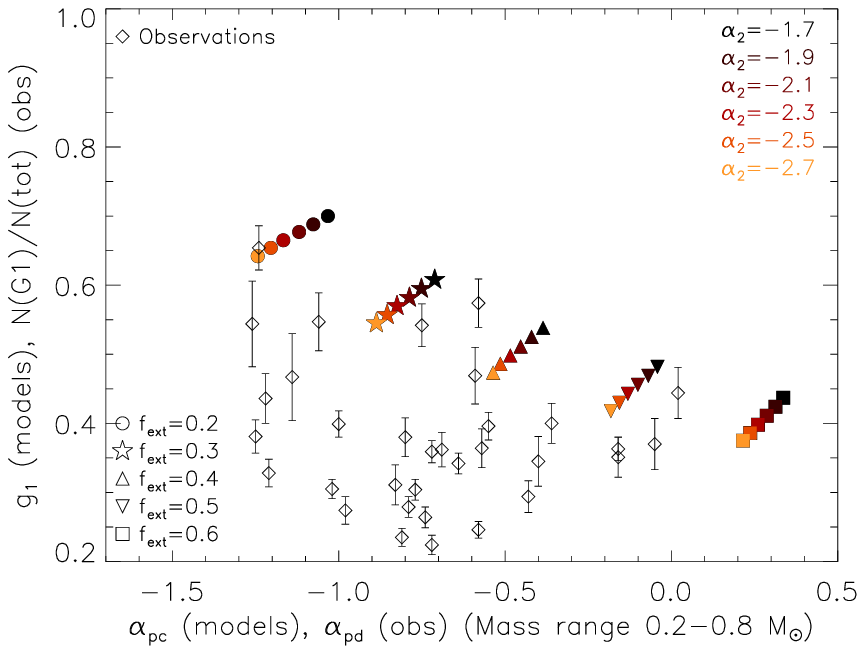}
\caption{Relationship between $g_{1}$ and the slope of the post-collision mass function, $\alpha_{pc}$ (for stars in the mass range 0.2-0.8 M$_{\odot}$), for various shapes of the IMF, namely for various values of the slope at the high mass range, $\alpha_{2}$, and for various values of $f_{ext}$. In all models, the values of $\alpha_{0}$ and $\alpha_{1}$ are fixed to $-0.3$ and $-1.3$, respectively. The models are compared to the $(N({\rm G1})/N({\rm tot}))-\alpha_{pd}$ relation found in the observations.}
\label{fig2}
\end{center}
\end{figure}
    
\section{Results}\label{res}

We describe the IMF of the G1 stars using a multicomponent power-law function, which is given by (Kroupa 2001) 

\begin{eqnarray}
\begin{array}{l} 
 \psi \left(M \right)=A 
 \\
 \end{array}
 \left\{
 \begin{array}{l}
\left(\frac{M} {0.08}\right)^{\alpha_{0}}, M \leq 0.08~{\rm M}_{\odot} \\ 
\left(\frac{M} {0.08}\right)^{\alpha_{1}}, 0.08~{\rm M}_{\odot} \leq M  \leq 0.5~{\rm M}_{\odot}\\
\left(\frac{0.5}{0.08}\right)^{\alpha_{1}}\left(\frac{M}{0.5}\right)^{\alpha_{2}}, M \geq 0.5~{\rm M}_{\odot},\\
\end{array}
\right.  
\label{eq7}
\end{eqnarray}and where $A$ is a normalization constant that depends on the cluster mass. Its determination is not necessary since we are only interested in the fractions of stars in the target and product mass bins. When calculating the fractions, we assumed the minimum and maximum stellar masses to be $M_{min}=0.01$ M$_{\odot}$ and $M_{max}=150$ M$_{\odot}$, respectively. In Eq.~\ref{eq7} the value of $\alpha_{0}$ is the most uncertain and displays a significant amount of scatter amongst the Milky Way clusters. We fixed its value to the one derived for the Galactic field, namely $\alpha_{0}=-0.3$. In the first instance, we considered the effect of varying the value of the slope at the high mass end, $\alpha_{2}$. The Galactic field value is $\approx -2.3$, and we considered a range of values going from $-1.7$ to $-2.7$. The value of $\alpha_{1}$ was set to the Galactic field value of $-1.3$. The shape of the IMF for various values of $\alpha_{2}$, for the same arbitrary mass reservoir, is displayed in Fig.~\ref{fig1}. For each set of free parameters ($\alpha_{2}$ and $f_{ext}$), we calculated the values $\alpha_{pc}$ and $g_{1}$ using the model described in Sect. \ref{mod}. The results in the $g_{1}-\alpha_{pc}$ space are displayed in Fig.~\ref{fig2} and are compared to the observational data. Variations in $\alpha_{2}$, coupled to variations in $f_{ext}$ in the range (0.2-0.6), can help explain a fraction of the scatter that is observed in the $(N({\rm G1})/N({\rm tot}))-\alpha_{pd}$ relation but fall short of explaining the values of $(N({\rm G1})/N({\rm tot}))$ that are $\lesssim 0.4$. 

\begin{figure}
\begin{center}
\includegraphics[width=\columnwidth] {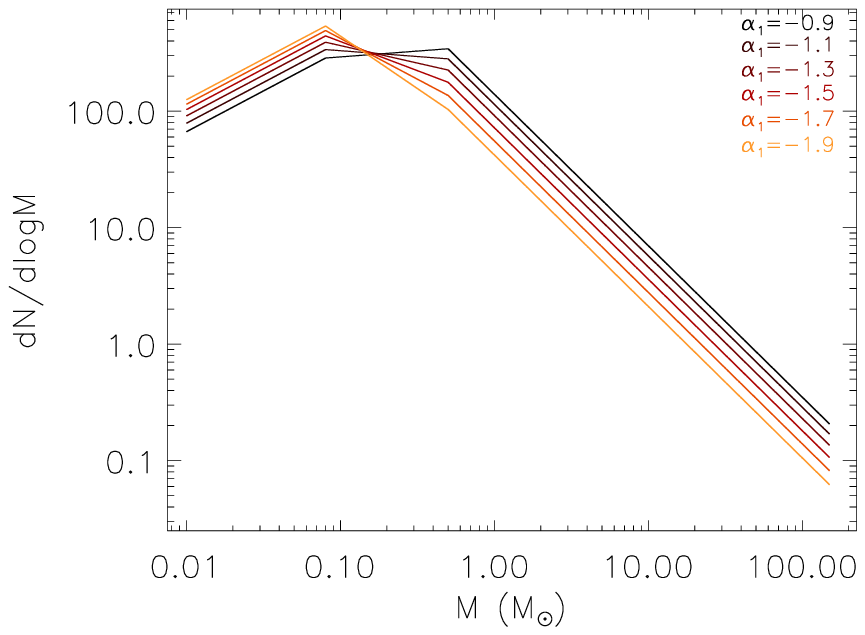}
\caption{Shape of the IMF for various values of the intermediate slope, $\alpha_{1}$, around the Galactic field value of $-1.3$. The values of $\alpha_{0}$ and $\alpha_{2}$ are set to the Galactic field values of $-0.3$ and $-2.3$, respectively. The mass functions are normalized assuming the same arbitrary mass.}
\label{fig3}
\end{center}
\end{figure}

\begin{figure}
\begin{center}
\includegraphics[width=\columnwidth] {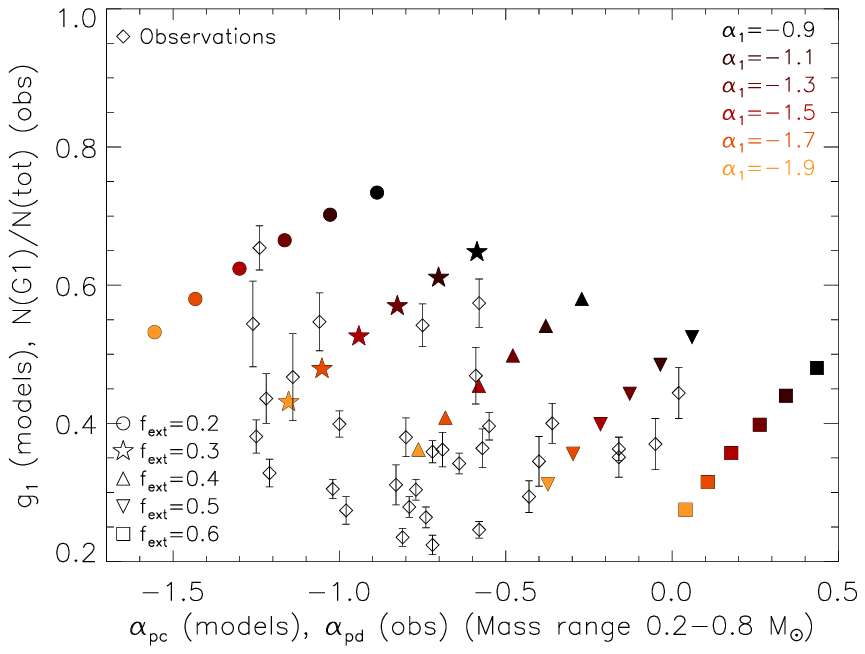}
\caption{Relationship between $g_{1}$ and the slope of the post-collision mass function, $\alpha_{pc}$ (for stars in the mass range 0.2-0.8 M$_{\odot}$), for various shapes of the IMF, namely for various values of the slope in the intermediate-mass range, $\alpha_{1}$, and for various values of $f_{ext}$. In all models, the values of $\alpha_{0}$ and $\alpha_{2}$ are fixed to $-0.3$ and $-2.3$, respectively. The models are compared to the $(N({\rm G1})/N({\rm tot}))-\alpha_{pd}$ found in the observations.}
\label{fig4}
\end{center}
\end{figure}

It is evident that variations in $(N({\rm G1})/N({\rm tot}))$ (equivalently in $g_{1}$ for the models) would be more significant if the primordial fractions of $G1$ stars were different, as a consequence of variations in the physical conditions of the star formation process. Dib (2014) and Dib et al. (2017) show that the characteristic mass (i.e., the peak in the IMF when fitted with a tapered power-law function) varies in the range $\approx 0.1$ to $\approx 0.8$ M$_{\odot}$ when measured for young clusters in the Milky Way. It is not clear whether the same level of variation is present in the IMF of GCs at the time they formed. Since we are using a broken power-law function with no well-defined peak, we varied the intermediate slope of the IMF in the mass range (0.08-0.5) M$_{\odot}$ (i.e., $\alpha_{1}$). Figure \ref{fig3} displays a number of IMF realizations for various values of $\alpha_{1}$ and where $\alpha_{0}$ and $\alpha_{2}$ are assigned the Galactic field values of $-0.3$ and $-2.3$, respectively. The derived $g_{1}-\alpha_{pc}$ relations for these different cases and for various values of $f_{ext}$ are displayed in Fig.~\ref{fig4}. As already shown in Kravtsov et al. (2022), varying the value of $f_{ext}$ for a fixed set of the IMF parameters reproduces the anticorrelation between $(N({\rm G1})/N({\rm tot}))-\alpha_{pd}$. However, in order to reproduce the observed level of scatter in the observations, it is necessary to allow, for a fixed value of $f_{ext}$, the existence of variations in the values of $\alpha_{1}$. This level of variation in $\alpha_{1}$ translates into a mean mass that varies between $0.2$ and $0.55$ M$_{\odot}$. This level of variation is consistent with what is observed for young stellar clusters in the Milky Way (Dib 2014; Dib et al. 2017). The simple models presented here reproduce the observations remarkably well. As discussed in Sect. \ref{dat}, the ratios of $(N({\rm G1})/N({\rm tot}))$ measured by Milone et al. (2017) are likely to systematically underestimate their true values. If that is indeed the case, it would bring our models into an even better agreement with the observations. It is also worth pointing out that variations in $\alpha_{1}$ and $\alpha_{2}$ can occur simultaneously. Furthermore, we have restricted the distinct power laws to mass ranges similar to those inferred for the Galactic field. There is, however, no indication that these limits hold for individual clusters; they may well vary from cluster to cluster, as is observed in young clusters in the Milky Way (Dib 2014). 

Dynamical interactions in the clusters that lead to the preferential ejection of low-mass stars can also cause the slope of the mass function in GCs to flatten with time  (e.g., Webb \& Vesperini 2016; Webb et al. 2017; Baumgardt \& Hilker 2018; Baumgardt et al. 2019). Sollima \& Baumgardt (2017) as well as Ebrahimi et al. (2020) plotted the value of $\alpha_{pd}$ as a function of the cluster age ($t_{age}$) expressed in units of the present-day half-mass relaxation time ($t_{rh}$). The positive correlation between $\alpha_{pd}$ and the ratio $(t_{age}/t_{rh}$) suggests that the slope becomes flatter with increasing values of ($t_{age}/t_{rh}$), and this effect can be understood in terms of the dynamical evolution of the clusters. De Marchi et al. (2010) fitted the present-day mass function of many Galactic clusters, young and old, with a tapered power-law function and deduced that the characteristic mass varies between 0.1 and 0.8 M$_{\odot}$, which is roughly the same range of variation implied by our model. However, De Marchi et al. attributed the difference in the characteristic mass to the preferential evaporation of low-mass stars from the clusters, which over time leads to a shift in the characteristic mass to higher values. 

\begin{figure}
\begin{center}
\includegraphics[width=0.9\columnwidth] {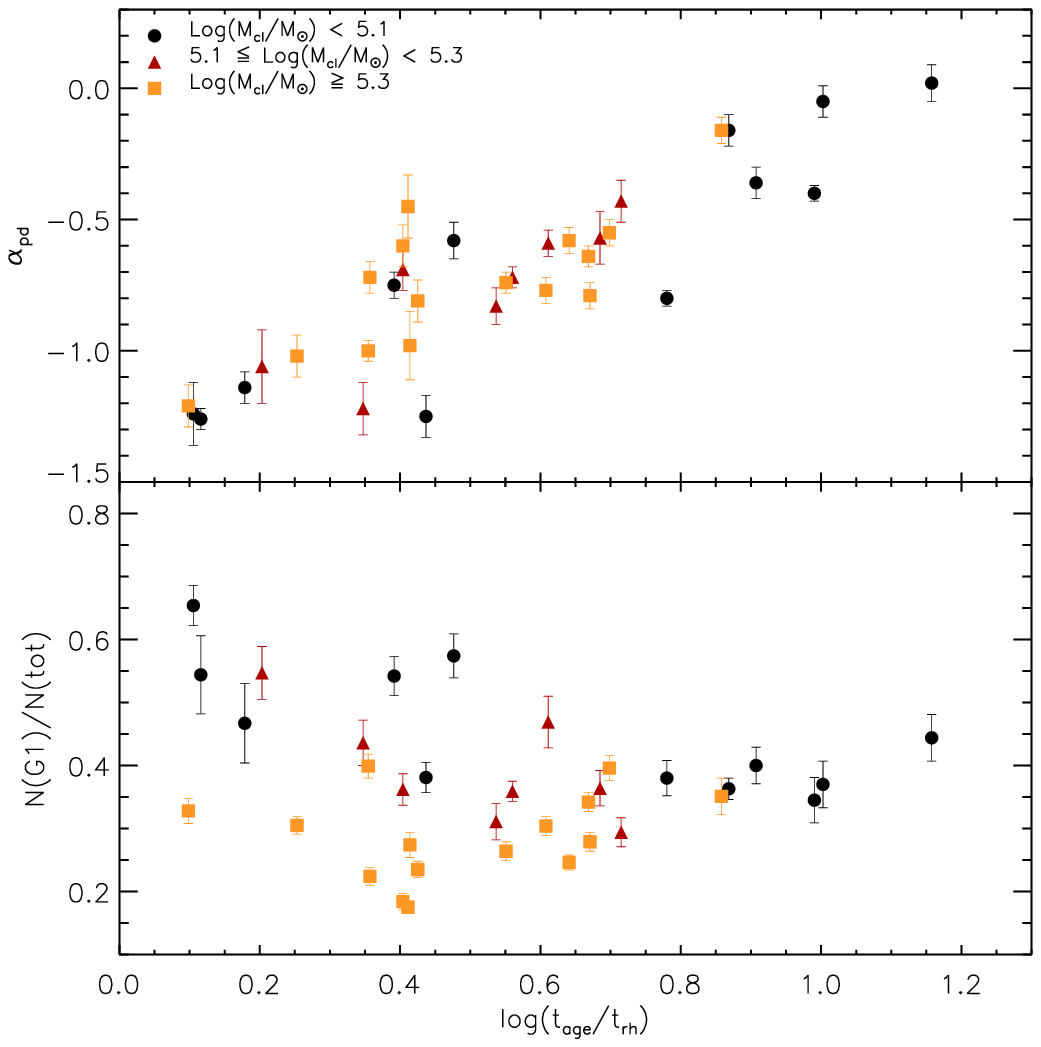}
\vspace{0.6cm}
\caption{Value of the present-day slope of the stellar mass function ($\alpha_{pd}$) plotted as a function of the ratio (${\rm log}(t_{age}/t_{rh})$), where $t_{age}$ are the ages of the clusters and the $t_{h}$ their present-day half-mass relaxation time (upper subpanel) and fraction of G1 stars in the clusters plotted as a function of ${\rm log}(t_{age}/t_{rh})$ (lower subpanel). In both subpanels, clusters are segregated by their mass, which is taken to be the mean value of the photometric and dynamical masses.}
\label{fig5}
\end{center}
\end{figure}

In the context of the formation of a second generation of stars, N-body models of GCs' evolution show that dynamical evolution preferentially removes low-mass stars from the clusters' centers, thus decreasing their fractions with time (e.g., Vesperini et al. 2013,2021; Sollima 2021). We note, however, that all of these N-body studies adopted a fixed form of the IMF (i.e., usually the Kroupa 2001 functional form) and did not allow for stellar collisions. In Fig.~\ref{fig5} (upper subpanel), we reproduce the $\alpha_{pd}$ versus ${\rm log}(t_{age}/t_{rh})$ plot of Ebrahimi et al. (2020), which is expanded by the inclusion of the additional clusters (see Sect. \ref{dat}) and where clusters are segregated by their mass, which we take to be the mean value between the photometric and dynamical mass. Figure~\ref{fig5} also displays the values of $(N({\rm G1})/N({\rm tot}))$ plotted as a function of ${\rm log}(t_{age}/t_{rh})$ (lower subpanel). Here again, the clusters are segregated by their mass. While it is possible to observe a weak anticorrelation between $(N({\rm G1})/N({\rm tot}))$ and ${\rm log}(t_{age}/t_{rh})$ for low-mass GCs in the sample (Spearman's $\rho$ coefficient $\approx -0.69$), this anticorrelation is not detected for high-mass clusters (Spearman's $\rho \approx 0.22$). It is possible that this "dichotomy" could indicate a stronger and weaker role for dynamical interactions in low- and high-mass clusters, respectively. This is confirmed by the results of Dalessandro et al. (2014), who find that G1 and G2 stars are entirely spatially mixed in the low-mass GC NGC 6362. Inversely, this would also indicate a more pronounced contribution from collisions in high-mass clusters and a weaker one in their low-mass counterparts. While it is probably too early to ascertain this fact from the current data, N-body models that follow the dynamical evolution of GCs while accounting for stellar collisions can help shed more light on this issue.

\section {Discussion and conclusions}\label{conc}

In this work we explore, in GCs that harbor multiple populations, the origin of the anticorrelation that is observed between the fraction of the first generation of stars ($N({\rm G1})/N({\rm tot})$) and the slope of the present-day mass function of the clusters in the stellar mass range (0.2-0.8) M$_{\odot}$ ($\alpha_{pd}$). We compare the observations to the results of a simple model that is based on the idea of stellar collisions between G1 stars in the mass range (0.1-0.5) M$_{\odot}$ that lead to the formation of a second generation of stars with masses in the range (0.5-0.9) M$_{\odot}$. The model has two input parameters, the shape of the IMF of the G1 stars and the fraction of G1 stars in the (0.1-0.5) M$_{\odot}$ mass range that collide to form G2 stars (parameter $f_{ext}$). The parameter $f_{ext}$ encapsulates much of the physics that governs the efficiency of the collision process. Its value depends on the compactness of the protocluster cloud, and hence on the mean distance between stars and also on other factors such as the initial levels of mass segregation in the clusters, the binary fraction, and the period distribution of binaries. We find that the appropriate range for $f_{ext}$ is $\approx (0.2-0.6)$. Smaller and larger values of $f_{ext}$ lead to values of the slope that are outside the range of observed values. 

Our results show that variations in $f_{ext}$ are necessary to explain the anticorrelation between $(N({\rm G1})/N({\rm tot}))$ and $\alpha_{pd}$ for a fixed shape of the IMF. However, the large scatter that is observed in the $(N({\rm G1})/N({\rm tot}))-\alpha_{pd}$ anticorrelation can only be explained, in the framework of this collision-based model, by variations in the IMF of the G1 stars. In particular, we show that variations in the slope of the IMF in the intermediate-mass regime of a Kroupa-like IMF ($\approx 0.08-0.5$) in the range $-1.9$ to $-0.9$ can reproduce the scatter in $(N({\rm G1})/N({\rm tot}))$ at a given value of $\alpha_{pd}$. This level of variation in $\alpha_{1}$ corresponds to a range of $\approx$(0.2-0.55) M$_{\odot}$ in the mean stellar mass in the clusters. In this work we have restricted the target G1 stars to the mass range (0.1-0.5) M$_{\odot}$. However, mergers between stars whose masses fall in the range (0.1-0.9) M$_{\odot}$ and that result in G2 stars with masses $\leq 0.9$ M$_{\odot}$ can further decrease the values of $g_{1}$ and bring the models into a better agreement with the observations. Variations in the IMF of GCs have been pointed out by other groups. Zonoozi et al. (2016) show that a dependence of the high mass end of the IMF on metallicity as proposed by Marks et al. (2012) can help explain the mass-to-light versus metallicity anticorrelation that is observed for the population of GCs in M31.  

While it is probably safe to say that the jury is still out concerning the relative importance of collisions versus dynamical interactions in modifying the shape of the IMF in GCs, the collision-based model presented here provides an explanation for the observed $(N({\rm G1})/N({\rm tot}))-\alpha_{pd}$ anticorrelation, and it is not yet clear if dynamical evolution can do the same. The level of variations in the IMF inferred in this work, particularly for the slope in the stellar mass range (0.08-0.5) M$_{\odot}$ that is needed in order to explain the scatter in the $(N({\rm G1})/N({\rm tot}))-\alpha_{pd}$ relation, is consistent with what is observed for young stellar clusters in the Milky Way (Dib 2014) and may reflect variations in the physical conditions of the GC parental clouds at the time the G1 population started to form, the occurrence of other processes such as gas accretion onto protostars (Dib et al. 2010), or collisions between protostars before they settle on the MS (Dib et al 2007). 

\begin{acknowledgements}

We thank the anonymous referee for comments and suggestions that helped improve the paper. We also thank Pavel Kroupa for useful exchanges on the issue of multiple populations in globular clusters.
 
 \end{acknowledgements}

{}

\label{lastpage}

\end{document}